\documentstyle[prl,aps,twocolumn,fleqn,epsfig]{revtex}
\begin{document}                
\draft
\title{Single ion anisotropy of Mn doped GaAs  measured by EPR.}
\author{O. M. Fedorych, E. M. Hankiewicz, Z. Wilamowski}
\address{Institute of Physics PAS, 02-668 Warsaw, Poland}
\author{J. Sadowski}
\address{MAX-Lab, Lund University, P. O. Box 118, SE-211 00 Lund, Sweden and Institute of Physics PAS, 02-668 Warsaw, Poland}
\address{
\centering{
\medskip\em
{}~\\
\begin{minipage}{14cm}{Electron paramagnetic resonance (EPR) study of MBE grown Mn doped GaAs
 is presented.  The resolved fine structure allows us to evaluate
 the crystal field parameters of the spin Hamiltonian. The obtained cubic
 constant is $a= -14.1\cdot 10^{-4}$ cm$^{-1}$. The axial field parameter, $D$, increases with
 Mn concentration, $x$, i.e., with the strain of Ga$_{1-x}$Mn$_{x}$As layers.  Extrapolation of
 $D$ shows that the single ion anisotropy is the important contribution to the magnetic
 anisotropy which is observed in ferromagnetic layers with greater Mn concentrations.
 The analysis of the EPR linewidth shows that native defects of the concentration of
 $5\cdot 10^{19}$ cm$^{-3}$, but not the Mn ions, are
 the main origin of crystal field fluctuations.}~\\ {}~
\end{minipage}
}}

\maketitle

    Ga$_{1-x}$Mn$_{x}$As is a semiconductor which exhibits ferromagnetic
properties \cite{tanaka,ohno1,ohno2,dietl}.  It is one of the most
promising diluted ferromagnets which can be applied in magnetic
semiconductor devices. The ferromagnetic phase occurs already for
$x$= 0.02 and for $x$= 0.04 the Curie temperature reaches 90 $K$.
Some logic structures, where the magnetic phase is switched on by
an electric gate have already been built \cite{koshihara}.

Ferromagnetic layers grown by low temperature molecular beam
epitaxy (MBE) exhibit an axial magnetic anisotropy which plays a
crucial role in magnetic properties of Ga$_{1-x}$Mn$_{x}$As layers
and multi layer structures.  The sign of the anisotropy depends on
the substrate.  For a tensile strain the direction parallel to the
growth direction is the easy axis while for compressive strain the
easy direction lies in the grown plane.  The magnitude of the
anisotropy field is of the order of 5-50 $mT$.  Definitely, the
anisotropy does not originate from magnetic dipole interactions.
Such a shape anisotropy should be by an order of magnitude smaller
and the easy axis would have to lie in plane.   The complex nature
of the valence band structure leads us to expect pseudo-dipole or
Dzialoschynski-Moriya interactions \cite{macdonald}.  They could
lead to a broadening of the resonance linewidth but their
theoretical estimations is not simple.  To distinguish whether the
anisotropy originates from a single ion anisotropy of Mn$^{2+}$
ion or from a non Heisenberg exchange coupling between Mn$^{2+}$
ions we have undertaken a detailed electron paramagnetic resonance
(EPR) study of Mn in MBE grown Ga$_{1-x}$Mn$_{x}$As.

Some reports on EPR of Ga$_{1-x}$Mn$_{x}$As were already published
\cite{almeleh,szczytko1,szczytko2}. The hyperfine structure (HFS)
is commonly observed but the fine structure (FS) of the EPR
spectrum has not been resolved until now. Almeleh and Goldstein
estimated the upper limit of the cubic component of the crystal
field parameter, $a$, in bulk samples \cite{almeleh}.  The axial
crystal field component, described by the parameter $D$, which may
occur in strained structures only, has not yet been analyzed.  In
this paper we present EPR studies of weakly doped
Ga$_{1-x}$Mn$_{x}$As compounds growth by MBE. We are able to find
a resolved FS and to evaluate the single ion anisotropy and
fluctuations of the crystal strain.  No broadening of EPR lines
caused by non-Heisenberg component of Mn-Mn coupling has been
found.

    The samples were grown by low temperature
molecular beam epitaxy (LT MBE) in a KRYOVAK MBE system
\cite{sad}. The calibration of Ga and Mn sources allows us two
estimate the Mn content with an accuracy better than 0.1\%.  The
composition of samples with low Mn concentrations was verified by
SIMS measurements and by the integrated amplitude of EPR signal.
The thickness of low concentration samples was about 2
micrometers. The XRD measurements showed that all samples with Mn
content from 10$^{-5}$ to 10$^{-3}$ are coherently strained by the
GaAs(100) substrate.

In EPR studies we found that samples with $x\leq1.5\cdot10^{-3}$
are characterized by a well resolved HFS and some weakly resolved
FS. For $x>2.5\cdot10^{-3}$ only a single broad line is observed.
The spectra of Ga$_{1-x}$Mn$_{x}$As ($x=4 \cdot 10^{-4}$) for four
various directions of the applied magnetic field are shown in Fig.
1.  The magnetic field was applied in the (110) crystal plane and
$\Theta$ is the angle measured from the [001] direction.

These spectra are characterized by $g=2$ and definitely originate
from $S=5/2$ of Mn$^{2+}$ ions.  Schneider et al. found additional
lines of a neutral acceptor with $g$=2.77 \cite{schneider}. We
observe only the signal of ionized acceptors which indicates that
LT MBE samples are strongly compensated.

\begin{figure}[tbh]
\mbox{}\centerline{\epsfig{file=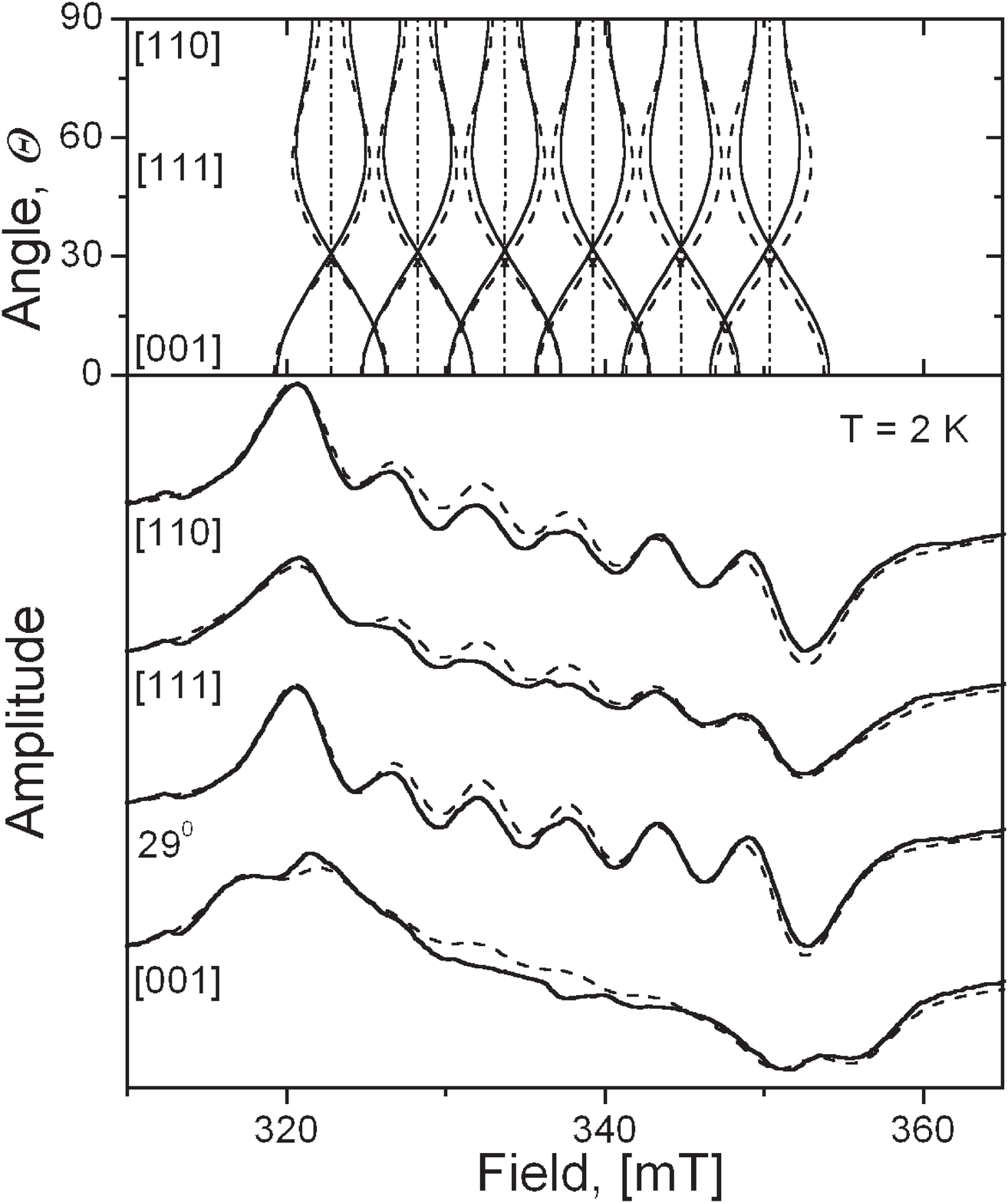,width=8cm}}
\caption{EPR spectra for various directions of external magnetic
field, H, applied in (110) plane.  The fine structure is well
resolved for H parallel  to [001] direction $( \Theta =0$ and $T
\leq 20$ K). The dashed line correspond to best fitted shapes. The
upper diagram shows the
 angular dependence of the EPR line positions.}
\end{figure}

The six lines of the HFS structure, corresponding to six possible
projections, $M_{I}$, of the $I=5/2$ nuclear spin of Mn, are well
resolved for an arbitrary direction of the magnetic field.  The
value of the HFS constant, $A= 51.6 \cdot 10^{-4}$ cm$^{-1}$,
agrees with that known from literature.  The dependence of the
spectrum on the magnetic field direction is caused by the angular
dependence of the FS structure.  The FS splitting is small and the
FS structure is resolved only for the magnetic field directed
along [001] ($\Theta=0$).  It manifests itself as extra lines
below the first and above the last of the six HFS lines.

We analyze the experimental data according to the standard spin
Hamiltonian of the form \cite{AB}:

\begin{eqnarray}\label{1}
 H _{S}= g \mu_{B}HS + ASI +D[S_{z}^{2}-(1/3)S(S+1)]+
 \nonumber\\
+(a/6)[S_{x}^{4}+S_{y}^{4}+S_{z}^{4}-(1/5)S(3S^{2}+3S+1)]
\end{eqnarray}
here the first term is the Zeeman interaction, $\mu_{B}$ is the
Bohr magneton.  The second term corresponds to the HFS coupling
which is assumed to be isotropic.  The third and the fourth terms
describe the FS structure, corresponding to the axial and the
cubic symmetry of the crystal field, respectively. The parameter
$D$ is directly related to the single ion contribution to the
magnetic anisotropy. The corresponding anisotropy field for
ferromagnetically saturated spins  is $H_{A} = S D/g \mu_{B}$.

The positions of 30 spectral lines, corresponding to six
orientations of the nuclear spin, $I$, ($M_{I}=-5/2,..., 5/2$) and
to five different spin flip transitions between six energy levels
labeled by $M_{S}$  ($M_{S}=-3/2,..., 5/2$) were calculated
according to spin Hamiltonian (1).  To evaluate the spin
Hamiltonian parameters we took the best fit of all spectra
measured for various field orientations $\Theta$.  Details of the
fitting procedure is discussed below.  The best fit was obtained
for $|a|=(14.1\pm0.3)\cdot10^{-4}$ cm$^{-1}$.  Within the
experimental error this value does not depend on the Mn
composition, $x$.  The value of the axial component of the crystal
field, $D$, is much smaller than the cubic parameter, $a$.  For
the data shown in Fig. 1, $D= (1.3\pm0.02) \cdot 10^{-4}$
cm$^{-1}$.  The dependence $D$ on $x$ is shown in Fig. 2.

Because the line broadening is mainly due to super HFS caused by
the coupling of Mn$^{2+}$ spins with the nuclear spins of the
ligands \cite{almeleh}, the shape of the individual spectral lines
is assumed to be described by a Gaussian function which gives a
much better fit as compared to fits when  Lorentz line shapes were
assumed. The satellite lines are characterized by a slightly
bigger linewidth.  In the fitting procedure we  used two
parameters describing the width of individual spectral lines: the
linewith of the central FS line, $\Delta H_{1/2,-1/2}$, and the
parameter describing an extra broadening of the satellite lines of
FS: $\delta \Delta H=\Delta H_{5/2,3/2} - \Delta H_{1/2,-1/2}$.
The width of the line (3/2,1/2) is assumed to be equal to the mean
value: $\Delta H_{3/2,1/2}=( \Delta H_{5/2,3/2}+ \Delta
H_{1/2,-1/2})/2$. The best fitted value of the linewidth is
$\Delta H_{1/2,-1/2}=(2.0 \pm 0.02)$ $mT$.  It is found to be
isotropic and independent of $x$.  The additional broadening  of
the satellite lines, $\delta \Delta H$, is a small effect.  For
the data shown in Fig. 1 $\delta \Delta H$=0.4 $mT$.  This
corresponds to a fluctuation of $D$, $\delta D = \delta \Delta H /
g \mu_{B}$, equal to $2\cdot 10^{-4}$ cm$^{-1}$.  As it is shown
in Fig. 2b this value varies from sample to sample by about 40\%
but there is no systematic variation of $\delta \Delta H$ with Mn
concentration.

The resolved FS and HFS vanish with the increase of Mn
concentration.  To simulate the observed shape of the EPR spectrum
it is not enough to consider the broadening of the individual
spectral lines.  A good fit is obtained only when we treat the
spectrum as a sum of the resolved one, originating from
individual, non interacting Mn$^{2+}$ spins, and the unresolved
contributions with averaged FS and HFS. With an increase of $x$
the width of individual lines of the resolved spectrum
($\Delta$$H_{1/2,-1/2}$ and $\delta \Delta H$) practically does
not change.  The resolved structure gradually disappears because
of the decrease of the amplitude of the resolved contribution. For
low $x$ the width of the unresolved contribution corresponds to
the second moment of the spectrum of non interacting Mn$^{2+}$
spins.  With a concentration increase a decrease of the width of
the unresolved spectrum is observed.  We relate this decrease to
the exchange narrowing effect.

\begin{figure}[tbh]
\mbox{}\centerline{\epsfig{file=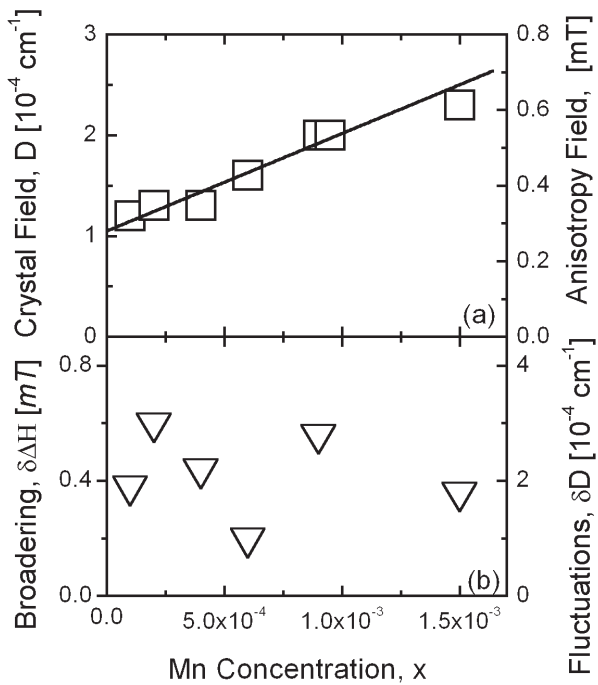,width=8cm}}
\caption{(a) The dependence of the single ion axial anisotropy $D$
on the Mn composition $x$, as a function of Mn composition, $x$.
The increase of $D$, $dD/dx = 0.09$ cm$^{-1}$, is caused by a
compressive strain originating from lattice mismatch.   The single
ion anisotropy fields $DS/g$$\mu$$_{B}$ are labeled on the right
axis.  (b) The difference of the linewidth of satellite and
central lines of FS, $\delta \Delta H$, scales fluctuations of the
crystal field, $\delta D$, caused by fluctuations of the
elementary cell.}
\end{figure}

The axial component of the crystal field anisotropy $D$ is caused
by a local strain acting on the ion.  It increases with an
increase of $x$.  The data shown in Fig. 2 can be approximated by
a linear dependence $D=(0.09 \cdot x+1.05 \cdot 10^{-4})$
cm$^{-1}$. The value $DS/g \mu _{B}$ corresponds to the single ion
contribution to the total anisotropy field (when a saturated
polarization of Mn spins is assumed).   For $x=1.5\cdot10^{-3}$
the anisotropy field is $DS/g \mu _{B}$= 0.7 $mT$.  However, since
the layer strain increases linearly with $x$ the single ion
anisotropy expected for $x=0.035$ is $DS/g \mu _{B}\approx 10$
$mT$. This value agrees well with the coercion field observed in
ferromagnetic samples \cite{ohno1,sad}.  It allows us to conclude
that the single ion anisotropy is the considerable contribution to
the total magnetic anisotropy of ferromagnetic layers.

The fact that the extrapolation of the parameter $D$ to low Mn
concentration,  $x \rightarrow 0$, does not  tend to zero we
relate to an additional strain which appears at low temperature
caused by a difference of the thermal expansion coefficients of
Ga$_{1-x}$Mn$_{x}$As, GaAs and sample holder.

The signs of the parameters $D$ and $a$ cannot be directly
determined from EPR measurement.  The data allow only to conclude
that the parameters are of opposite sign, $a \cdot D<0$.  However,
from the fact that the magnetic anisotropy in ferromagnetic
Ga$_{1-x}$Mn$_{x}$As layers grown on GaAs substrate shows an easy
in-plane magnetization, we conclude that $D>0$, and, in
consequence, we suggest that $a<0$.

Within the whole range of composition where the EPR spectrum is
well resolved the linewidth of the central FS line, $\Delta
H_{1/2}$, is constant.  Up to $x=1.5\cdot10^{-3}$ we do not
observe any additional broadening of the resolved part of the
spectrum which could be attributed to  dipole or pseudo-dipole
(non-Heisenberg exchange coupling) interactions.  The resolved
part of the EPR spectrum, however, does not originate from a
statistic Mn spin but from those Mn ions which do not have any
other Mn ion in the neighborhood.   Because of that we are not
able to conclude about the magnitude of a mean dipole or
pseudo-dipole coupling between Mn ions.  Moreover, the discussed
samples are strongly compensated. Because of that the fact that no
effect of non-Heisenberg Mn-Mn coupling has been found does not
allow us to made any solid conclusion about the character of RKKY
interactions.

The additional broadening of the satellite lines of FS, $\delta
\Delta H$, can be attributed to fluctuations of the local strain.
Because the position of the central FS line is not affected by
crystal field, the strain induced fluctuations of FS structure
parameters, $a$ and $D$ (including a random direction of the local
strain), do not affect the width of the central FS line, $\Delta$
$H_{1/2,-1/2}$, but they lead to a broadening of the satellite
lines only.  Consequently, $\delta \Delta H$ scales the amplitude
of the strain fluctuation.  Assuming that the evaluated derivative
$dD/dx$=0.09 cm$^{-1}$ (see Fig. 2) is directly related to the
difference of the lattice constants between Ga$_{1-x}$Mn$_{x}$As
and the GaAs substrate \cite{ohno1,sad} one can estimate
fluctuations of the deformation of the crystal cell, $a _{\circ}$,
in the vicinity of Mn dopant caused by other impurities to be
$\delta a _{ \circ}=10^{-4} a_{\circ}$. This value is considerably
bigger than the mean fluctuations amplitude as measured by rocking
curve of XRD. The fact that $\delta a _{ \circ}$ is independent of
$x$ indicates that the fluctuations, $\delta a _{ \circ}$, do not
originate from neighboring Mn ions but from other, non magnetic
defects, probably antisites.

Concluding, the single ion anisotropy caused by the layer strain
is an important  contribution to the whole magnetic anisotropy
observed in ferromagnetic Ga$_{1-x}$Mn$_{x}$As layers.  Discussion
of the local strain, as measured by broadening of the satellite
lines of FS, $\delta \Delta H$, allows to estimate crystal strain
fluctuations. The independence of the crystal field
fluctuations, 
of $x$ indicates that the concentration of native defects is
independent of the Mn content.  Assuming that the range of $x$
where the well resolved EPR spectrum of ionized Mn acceptors is
observed corresponds to the range of insulating and fully
compensated crystal, the upper limit of this range, $x=2.5 \cdot
10^{-3}$, allows to estimate the concentration of compensating
centers to be $5 \cdot 10 ^{19}$ cm$^{-3}$.

{\bf Acknowledgements}:Valuable discussion with H. Przybylinska
and F. Matsukura.  Work supported by the KBN grant 2 P03B 007 16.


\begin{references}
 \bibitem{tanaka} M. Tanaka, I. Vac.Sci.Technol. B {\bf16}, 2267 (1998)
 \bibitem{ohno1} H. Ohno, J. Magn. Magn. Mat. {\bf200}, 110 (1999)
 \bibitem{ohno2} Y. Ohno, D.K. Young, B. Beschoten, F.Matsukura, H.Ohno, and D. D. Awschalom, Nature, {\bf402}, 790 (1999)
 \bibitem{dietl} T. Dietl, H. Ohno,  F. Matsukura, J. Cibert, and D. Ferrand, Science, {\bf287}, 1019 (2000)
 \bibitem{koshihara} H. Ohno, D. Chiba, F. Matsukura, T. Omiya, E.
 Abe, T. Dietl, Y. Ohno, and K. Othani, Nature {\bf408}, 944 (2000)
 \bibitem{macdonald} M. Abolfath, T. Jungwirth, J. Brum, and A.H. MacDonald, Phys. Rev. B {\bf63}, 54418 (2001)
 \bibitem{almeleh} N. Almeleh and B. Goldstein, Phys. Rev. {\bf128}, 1568 (1962)
 \bibitem{szczytko1} J. Szczytko, W. Mac, A. Twardowski, F. Matsukura, and H. Ohno,Phys. Rev. B, {\bf59}, 12935 (1999)
 \bibitem{szczytko2} J. Szczytko, A. Twardowski, K. \'{S}wi\c{a}tek, M. Palczewska, M. Tanaka, T. Hayashi,K. Ando, Phys. Rev. B {\bf60}, 8304 (1999)
 \bibitem{sad} J. Sadowski, J. Z. Domaga\l, J. B\c{a}k - Misiuk, S.Kole\'{s}nik, M. Sawicki, K. \'{S}wi\c{a}tek, J. Kanski, L. Ilver, V. Str\"{o}m, J. Vac. Sci. Technol. B, {\bf18}, 1697 (2000)
 \bibitem{schneider} J. Schneider, U. Kaufmann, W. Wilkening, M. Baeumler and F. K\"{o}hl, Phys. Rev. Lett. {\bf59}, 240, (1987).
 \bibitem{AB} A. Abragam and B. Bleaney, Electron Paramagnetic Resonance of Transition Ions (Claredon, Oxford, 1970).
 \end{references}
\end{document}